\documentclass[a4j,11pt]{article}

\usepackage{amsmath}
\usepackage{amsmath,amssymb,bm}
\usepackage[dvipdfm]{graphicx, color}

\def\beq {\begin{equation}}
\def\eeq {\end{equation}}
\def\beqn {\begin{eqnarray}}
\def\eeqn {\end{eqnarray}}
\def\bmat {\begin{pmatrix}}
\def\emat {\end{pmatrix}}

\title{Neutrino properties in $E_6\times SU(2)_F$ SUSY GUT\\ with spontaneous CP violation}

\author{
\centerline{
Nobuhiro~Maekawa$^{1,2}$\footnote{E-mail address: maekawa@eken.phys.nagoya-u.ac.jp}
~and 
Kenichi~Takayama$^{1}$\footnote{E-mail address: takayama@th.phys.nagoya-u.ac.jp}}
\\*[25pt]
\centerline{
\begin{minipage}{\linewidth}
\begin{center}
$^1${\it \normalsize Department of Physics, Nagoya University, Nagoya 464-8602, Japan }  \\*[10pt]
$^2${\it \normalsize Kobayashi Maskawa Institute, Nagoya University, Nagoya 464-8602, Japan }  \\*[10pt]\end{center}
\end{minipage}}
\\*[50pt]}

\date{}

\begin{document}

\maketitle

\begin{abstract}

We examined the neutrino sector in $E_6\times SU(2)_F$ SUSY GUT with spontaneous CP violation.
At a glance, the discrete symmetry, which is introduced in order to solve the SUSY CP problem, constrains the allowed operators too strongly for the neutrino sector to be consistent with the experimental data, i.e., the $\mu$ neutrino becomes massless as commented in the previous paper.
We showed that this issue can be solved if some operators are taken into account.
And we saw that such operators do not play an important role in studying quark and charged lepton sectors.
The predictions on the neutrino masses and mixings are the same as the $E_6$ models, which are consistent with various experiments on the neutrino oscillations.

\end{abstract}
\vspace{1cm}

\section{Introduction}

Weak scale supersymmetry (SUSY) is one of the most promising candidates for physics beyond the Standard Model (SM) \cite{Nilles:1983ge}.
The minimal SUSY extension of the Standard Model (MSSM) has several attractive features.
It, for instance, provides a solution to the gauge hierarchy problem and a dark matter 
candidate as the lightest supersymmetric particle (LSP).
Moreover, supersymmetric grand unified theory (SUSY GUT) is strongly motivated by remarkable coincidence of three SM gauge coupling constants around $10^{16}$ GeV.

Grand unified theory can also explain the origin of hierarchical structures of masses and mixing angles in the SM particles by simply assuming that the $\bm{10}$'s of $SU(5)$ induce stronger hierarchical Yukawa structure than the $\bf{\bar 5}$'s.
However, the SUSY GUT scenario is generically suffering from various theoretical and phenomenological difficulties.
For instance,
\begin{enumerate}
\item SUSY flavor problem: 
if generic soft SUSY breaking terms are introduced, flavor changing neutral current (FCNC) processes exceeds the experimental constraints \cite{Gabbiani:1996hi}.
\item SUSY CP problem: 
if generic soft SUSY breaking terms are introduced, CP violating observables such as electric dipole moments (EDMs) exceeds the experimental constraints \cite{Gabbiani:1996hi}.
\item $\mu$ problem: 
supersymmetric Higgs mass $\mu$ must be the same order of the soft SUSY breaking scale though it is a SUSY parameter.
\item The doublet-triplet splitting (DTS) problem and proton decay: 
there must be huge mass separation between doublet Higgs (weak scale) and triplet Higgs (above GUT scale) in order to make proton's lifetime longer than experimental limit.
Moreover, recent proton decay constraint suggests that triplet Higgs mass should naively be larger than $10^{18}$ GeV, but it generically spoils the success of the gauge coupling
 unification.
\item Unrealistic GUT relation: 
generically, unification of quark and leptons in GUT tends to result in unrealistic Yukawa relations which are inconsistent with the observed masses and mixings of quark and leptons.
\end{enumerate}

$E_6$ unification is quite attractive because it can naturally induce the feature that
the $\bm{10}$'s of $SU(5)$ induce stronger hierarchical Yukawa structure than the $\bf{\bar 5}$'s, which plays an important role in obtaining realistic Yukawa hierarchies in $SU(5)$ GUT.
Moreover, if we introduce a family symmetry, $SU(3)_F$($SU(2)_F$), all three generation quark
and leptons can be unified into a single (two) multiplet(s), and after breaking the family
symmetry and the $E_6$ symmetry, realistic quark and lepton masses and mixings can be 
obtained. 
Furthermore the third generation $\bm{10}$ of $SU(5)$ can have different sfermion masses from
the other sfermion masses, because it is from the third generation $\bm{27}$
of $E_6$. 
It is remarkable that such effective SUSY sfermion mass spectrum \cite{Dimopoulos:1995mi} can satisfy the LHC
constraints with lighter stop, which is important to keep the naturalness in SUSY models.

Unfortunately, such effective SUSY type sfermion mass spectrum is generically suffering from
the new type of SUSY CP problem, in which Chromo electric dipole moment (CEDM) of up quark
becomes too large to satisfy the experimental constraints. The 1-3 mixings of up-type squark
masses become too large and have generically ${\cal O}(1)$ complex phase after diagonalizing
the complex Yukawa couplings if the stop mass is different from the other two up-type squarks.

Recently a scenario which solves this problem as well as the old type SUSY CP problem
by introducing spontaneous CP violation has been proposed \cite{Ishiduki:2009vr}.
The essential point is that the discrete symmetry, which is introduced in order to
solve the old type SUSY CP problem, leads to real up-type Yukawa couplings which
can solve the new type SUSY CP problem and to complex down-type Yukawa couplings which
give ${\cal O}(1)$ Kobayashi-Maskawa (KM) \cite{Kobayashi:1973fv} phase.
The model has remarkable features, for example,
the doublet-triplet splitting is naturally realized by introducing the anomalous $U(1)_A$
gauge symmetry, predictive Yukawa matrices of up quarks, down quarks and charged leptons are obtained
because of the reduced number of the ${\cal O}(1)$ parameters, smaller Cabibbo-Kobayashi-Maskawa (CKM) \cite{Cabibbo:1963yz} matrix element $V_{ub}\sim\lambda^4$, where $\lambda\sim 0.22$ is the Cabibbo mixing, is obtained.

In this paper, we examine the neutrino sector in the above model with the spontaneous CP
violation, since this has not been studied in detail in the previous paper. 
At a glance, the discrete symmetry, which is introduced in order to solve the SUSY CP problem,
constrains the allowed operators too strongly for the neutrino sector to be consistent with
the experimental data, i.e., the $\mu$ neutrino becomes massless as commented in the 
previous paper.
We will show that this issue on the neutrino sector can be solved if some operators, which are not important in studying quark and charged lepton sectors, are taken into
account. And we clarify the reason why such operators play an important role only in the
neutrino sector. The predictions on the neutrino masses and mixings are the same as the
$E_6$ models, i.e., $\sin\theta_{12}\sim\sin\theta_{23}\sim\lambda^{0.5}$, 
$\sin\theta_{13}\sim\lambda$, $\Delta m^2_{12}/\Delta m^2_{23}\sim \lambda^{2}$,
etc, which are consistent with various experiments on the neutrino oscillations.
Since the operators are restricted by the discrete symmetry, the predictions on the ${\cal O}(1)$
coefficients are expected. However, we are not able to find any simple predictions between
the ${\cal O}(1)$ coefficients because there are a lot of operators for the right-handed neutrino
masses.

As the result, we can obtain an attractive model in which 
\begin{enumerate}
\item realistic quark and lepton masses 
and mixings are obtained, 
\item all the CP phases in the model,
the KM phase, CP phases in neutrino sector and CP phases in sfermion sectors, 
are given from one phase which is obtained by the spontaneous CP violation,
\item as the result, SUSY CP problem is solved,
\item the effective SUSY type sfermion mass spectrum is predicted.
\end{enumerate}

The paper proceeds as follows: in Section 2, we briefly review the basic features of $E_6\times SU(2)_F\times U(1)_A$ SUSY GUT with spontaneous CP violation.
In Section 3, we calculate neutrino masses and Maki-Nakagawa-Sakata (MNS) \cite{Maki:1962mu} matrix explicitly, after specifying massless $\bar{\bm{5}}$ modes.
We see that a specific type of higher dimensional operators is essential for the realistic neutrino masses.
We also show that the above higher dimensional operators only affect the neutrino sector.
The last section is devoted to summary and discussion.

\section{Basic features of $E_6\times SU(2)_F\times U(1)_A$ SUSY GUT with spontaneous CP violation}

\subsection{$E_6$ unification and $SU(2)_F$ family symmetry}

First of all, $\bm{27}$ is the fundamental representation for the $E_6$ group.
In terms of $E_6\supset SO(10)\times U(1)_{V'}$ (and [$SO(10)\supset SU(5)\times U(1)_V$]) it is decomposed as
\begin{equation}
\bm{27}=\bm{16}_1[\bm{10}_{1}+\bar{\bm{5}}_{-3}+\bm{1}_{5}]+\bm{10}_{-2}[\bm{5}_{-2}+\bar{\bm{5}}'_{2}]+\bm{1}'_4[\bm{1}'_0],
\end{equation}
where acutes are used to distinguish different $\bar{\bm{5}}(\bm{1})$'s.
Note that each $\bm{27}$ contains two $\bar{\bm{5}}$'s (and $\bm{1}$'s) of $SU(5)$.
This nature is essential for realizing different Yukawa structures of up-type quarks, down-type quarks, charged leptons and neutrinos from a single hierarchical structure of an $E_6$ invariant Yukawa couplings \cite{Bando:2001bj, Bando:1999km, Maekawa:2002eh}.

In order to break $E_6$ gauge group into the SM gauge group $G_{\rm{SM}}$, we introduce three types of Higgs fields.
First one is $\bm{27}$ representation Higgs $H$.
We also introduce $\bm{\bar{27}}$ representation Higgs $\bar{H}$ in order to satisfy the
 D-flatness condition of $E_6$ gauge interaction.
$H, \bar{H}$ obtain vacuum expectation values (VEVs)
\begin{equation}
\langle H\rangle=\langle \bm{1}_H\rangle\neq 0,
\end{equation}
and they break $E_6$ into $SO(10)$.
Second one is an adjoint Higgs $A$, which belongs $\bm{78}$ representation.
It gets Dimopoulos-Wilczek (DW) \cite{Srednicki:1982aj} VEV proportional to $U(1)_{B-L}$ direction
\begin{equation}
\langle A\rangle=\langle \bm{45}_A\rangle \propto U(1)_{B-L},
\end{equation}
and it breaks $SO(10)$ into $SU(3)_C\times SU(2)_L\times SU(2)_R\times U(1)_{B-L}$.
Third type are $C$ and $\bar{C}$, and they are also $\bm{27}$ and $\bm{\bar{27}}$ representations.
They acquire VEVs
\begin{equation}
\langle C\rangle=\langle \bm{16}_C[\ni \bm{1}_C]\rangle\neq 0,
\end{equation}
and they break $SU(3)_C\times SU(2)_L\times SU(2)_R\times U(1)_{B-L}$ into $G_{\rm{SM}}$.
The superpotential which gives superheavy masses for ($\bm{5}, \bar{\bm{5}}$) and ($\bm{5}, \bar{\bm{5}}'$) pairs is
\begin{equation}
W=Y^H_{ij}\Psi_i\Psi_j H+Y^C_{ij}\Psi_i\Psi_j C,
\end{equation}
where $\Psi_i\ (i=1,2,3)$ are matter fields in family $i$.
After $H$ and $C$ acquire VEVs, three degrees of freedom among $\bar{\bm{5}}_i$ and $\bar{\bm{5}}'_i$ get superheavy masses, while remaining three are massless at the GUT scale.
We assume the Yukawa hierarches
\begin{equation}
Y^H_{ij}\sim Y^C_{ij}\sim \left(
\begin{array}{ccc}
\lambda^6 & \lambda^5 & \lambda^3\\
\lambda^5 & \lambda^4 & \lambda^2\\
\lambda^3 & \lambda^2 & 1
\end{array}\right)
\label{eq:Yukawa_hierarchy}
\end{equation}
and ratio of two VEVs
\begin{equation}
\frac{\langle C\rangle}{\langle H\rangle}\sim \lambda^{0.5}
\label{eq:VEV_ratio}
\end{equation}
up to ${\cal O}(1)$ coefficients.
These assumption can be realized proper charge assignment of anomalous $U(1)_A$ symmetry and the breaking effect of $SU(2)_F$ family symmetry, which we explain later respectively\footnote{Strictly speaking, (\ref{eq:VEV_ratio}) is not true in $U(1)_A$ framework.
The correct relation is $\lambda^c\langle C\rangle/\lambda^h\langle H\rangle\sim \lambda^{0.5}$, where $c$ and $h$ are $U(1)_A$ charges of $C$ and $H$, respectively.}.
Under these assumption, three massless modes are mainly ($\bar{\bm{5}}_1, \bar{\bm{5}}'_1, \bar{\bm{5}}_2$).
Note that third generation does not appear at the leading order.
So we end up following different Yukawa hierarchies
\begin{equation}
Y_d\sim \left(
\begin{array}{ccc}
\lambda^6 & \lambda^5 & \lambda^3\\
\lambda^5 & \lambda^4 & \lambda^2\\
\lambda^4 & \lambda^2 & 1
\end{array}\right), \ 
Y_d\sim Y_e^T\sim \left(
\begin{array}{ccc}
\lambda^6 & \lambda^{5.5} & \lambda^5\\
\lambda^5 & \lambda^{4.5} & \lambda^4\\
\lambda^3 & \lambda^{2.5} & \lambda^2
\end{array}\right), \ 
M_\nu\sim \lambda^n\left(
\begin{array}{ccc}
\lambda^7 & \lambda^{6.5} & \lambda^6\\
\lambda^{6.5} & \lambda^6 & \lambda^{5.5}\\
\lambda^6 & \lambda^{5.5} & \lambda^5
\end{array}\right)\frac{\langle H_u\rangle^2}{\langle \bar{H}\rangle^2}\Lambda,
\end{equation}
when the MSSM Higgses $H_u$ and $H_d$ are included in $\bm{10}_H$ and $\bm{16}_C$.
Here $n$ is a number, and we used the relation $\langle\bar{C}\rangle / \langle\bar{H}\rangle\sim \lambda^{0.5}$.

In order to realize the Yukawa hierarchy between ${\cal O}(1)$ top quark and the other quarks naturally, we employ $SU(2)_F$ family gauge symmetry.
We treat first two generation fields as the doublet $\Psi_a\ (a=1,2)$ whereas the third generation fields $\Psi_3$ and all Higgs fields are treated as singlets under $SU(2)_F$.
Here the index $a$ can be raised or lowered by the anti-symmetric symbols $\epsilon^{ab}$ and $\epsilon_{ab}$.
We also introduce flavon fields $F_a$ and $\bar{F}^a$ that are singlets under $E_6$ and doublet and anti-doublet under $SU(2)_F$, respectively.
VEVs of $F_a$ and $\bar{F}^a$ break $SU(2)_F$, then this effect generate hierarchical Yukawa structure in (\ref{eq:Yukawa_hierarchy}).
Since massless $\bar{\bm{5}}$'s do not contain the third generation at the leading order in $E_6$ GUT, both left-handed and right-handed components of the top quark belongs $SU(2)_F$ singlet, while all other quarks and charged leptons are (partially or completely) $SU(2)_F$ doublets.

$SU(2)_F$ family symmetry is also useful for solving the SUSY FCNC problem.
Suppose that the soft SUSY breaking terms are mediated to the visible sector above the scale where $E_6$ and $SU(2)_F$ symmetries are respected, such as in gravity mediation.
Then the symmetry guarantees  sfermion masses which degenerate at the leading order except for $\bm{10}_3$ \cite{Maekawa:2002eh}: 
\begin{equation}
\tilde m_{\bm{10}}^2=\left(\begin{array}{ccc}
m_0^2 & 0 & 0 \\
0 & m_0^2 & 0 \\
0 & 0 & m_3^2
\end{array}\right), \quad\quad
\tilde m_{\bm{\bar{5}}}^2=\left(\begin{array}{ccc}
m_0^2 & 0 & 0 \\
0 & m_0^2 & 0 \\
0 & 0 & m_0^2
\end{array}\right),
\label{eq:MUSM}
\end{equation}
where $\tilde{m}_{\bm{10}}$ and $\tilde{m}_{\bar{\bm{5}}}$ are squared sfermion mass matrices at the GUT scale in $\bm{10}$ and $\bar{\bm{5}}$ fields of $SU(5)$, respectively.
For sfermions in $\bm{10}$, FCNC observables provide stringent constraints mainly on the first two generations, while constraints for $\bar{\bm{5}}$'s are stringent for all three generations.
This is because the mixing angles of $\bm{10}$ fields are small, while those of $\bar{\bm{5}}$ fields are large.
Therefore, in this model, FCNC constraints can be evaded by raising $m_0$\footnote{There is upper limit of $m_0$ comes from charge and color breaking (CCB) effect.
For detail discussion, see \cite{ArkaniHamed:1997ab}.}.
Moreover, the weak scale is not destabilized as long as $m_3$ is around the weak scale.
So this type of sfermion masses (`` effective SUSY" or ``{\it modified} universality")
 \cite{Maekawa:2002eh, Kim:2006ab, Kim:2008yta, Ishiduki:2009gr, Kim:2009nq} solves the SUSY FCNC problem without spoiling naturalness.
We also comment that LHC constraints to this type of sfermion masses is much weaker than 
to the constrained MSSM (CMSSM) type sfermion masses \cite{Sakurai:2011pt}.

Unfortunately, there is a tension between this modified universal sfermion masses and
complex Yukawa couplings which are important to obtain the non-vanishing KM phase. In the basis in which Yukawa matrices are diagonal, sfermion mass matrices become
\begin{equation}
\Delta_{\bf 10}\equiv V_{\bf 10}^\dagger \tilde m_{\bf 10}^2 V_{\bf 10}
\sim \tilde{m}^2_{\bm{10}}+(m_3^2-m_0^2)\left(\begin{array}{ccc}
\lambda^6 & \lambda^5 & \lambda^3 \\
\lambda^5 & \lambda^4 & \lambda^2 \\
\lambda^3 & \lambda^2 & 1
\end{array}\right),  
\Delta_{\bf\bar 5}\equiv V_{\bf \bar 5}^\dagger \tilde m_{\bf\bar 5}^2 V_{\bf\bar 5}=
\tilde m_{\bf\bar 5}^2,
\end{equation}
where $V_{\bf 10}$ and $V_{\bf\bar 5}$ are the diagonalizing unitary matrices for 
$\bm{10}$ and $\bar{\bm{5}}$ fields of $SU(5)$, respectively, and we take
the CKM like matrix as $V_{\bf 10}$. Since the mass
insertion matrices $\delta_{\bf 10}\equiv \Delta_{\bf 10}/m_0^2$ do not vanish in the
limit $m_0\rightarrow\infty$, the CEDM of up quark from this SUSY contribution are
not decoupled. The constraints for these parameters by mercury (neutron) become \cite{Ishiduki:2009gr, Griffith:2009zz, Baker:2006ts, Hisano:2004tf}
\begin{equation}
{\rm Im}[(\delta_{u_L})_{13}(\delta_{u_R})_{31}]\leq 3\times 10^{-7}(9\times 10^{-7})
\left(\frac{m_3}{500 {\rm GeV}}\right)^2,
\end{equation}
which are much smaller than the prediction $\lambda^6\sim 10^{-4}$. 
This is a serious problem on the ``effective SUSY" or ``modified universality", which
we call the new type of SUSY CP problem.

\subsection{Anomalous $U(1)_A$ symmetry}

An Anomalous $U(1)_A$ symmetry \cite{Witten:1984dg, Froggatt:1978nt} is introduced in order to solve the doublet-triplet splitting problem and $\mu$ problem, and to provide the origin of the hierarchical Yukawa structures \cite{Maekawa:2001uk}.
This is a gauge symmetry whose anomalies are canceled by the Green-Schwarz mechanism \cite{Green:1984sg}.
The theory possesses the Fayet-Iliopoulos term $\xi^2$, and we assume its magnitude as $\xi=\lambda\Lambda$.
Here $\Lambda$ is the cutoff scale of the theory and we set $\Lambda=1$.
Let us denote the gauge symmetries of the theory except for $U(1)_A$ as $G_a$.
Consider a theory consisting of all the $G_a$ and $U(1)_A$ invariant terms including non-renormalizable operators.
Then it is shown in \cite{Maekawa:2001uk} that the theory has a supersymmetric vacuum where all the fields that are negatively charged under $U(1)_A$ get VEVs in the following way\footnote{From now on, each superfield is denoted by an uppercase letter, whereas the corresponding lowercase letter indicates an associated $U(1)_A$ charge.
The consistency of (\ref{eq:U(1)A_VEV}) requires the number of positively charged fields to be larger than that of negatively charged fields by one.}.
\begin{equation}
\left\{\begin{array}{cc}\langle Z_i^+\rangle = 0\ \quad & (z_i^+>0)\\
\langle Z_i^-\rangle \sim \lambda^{-z_i} & (z_i^-<0)\end{array}\right.
\label{eq:U(1)A_VEV}
\end{equation}
Here $Z_i$ is $G_a$ singlet field whose $U(1)_A$ charge is $z_i$.
This argument can be extended to the case where $Z_i$ is composite operator that is made by $G_a$ nonsinglet fields.
For example, in $Z^-=\bar{X}X$ case, $\langle \bar{X}X\rangle \sim \lambda^{-(x+\bar{x})}$ leads to $|\langle X\rangle| = |\langle\bar{X}\rangle| \sim \lambda^{-(x+\bar{x})/2}$, once the D-flatness condition of $G_a$ is taken account. Note that the above results can be 
obtained under a natural assumption that all the interactions which are allowed by the 
symmetry of the theory are introduced with ${\cal O}(1)$ coefficients.

It is important to mention that a term whose total $U(1)_A$ charge is negative does not appear at the $U(1)_A$ breaking vacuum.
The reason is that this type of term should originally be accompanied by at least one positively charged field but its VEV is always vanishing according to (\ref{eq:U(1)A_VEV}) (SUSY-zero mechanism) \cite{Maekawa:2001uk, Nir:1993mx}.
A $U(1)_A$ symmetry and its specific SUSY vacuum are applied in several aspects of phenomenological model building.
For example, an appropriate $U(1)_A$ charge assignment for the Higgs sector can ensure the DTS via the DW mechanism \cite{Maekawa:2001uk}, and the SUSY zero mechanism can be applied to solve the $\mu$ problem \cite{Maekawa:2001yh}.

For the following arguments, we will briefly remind the points on the solution for the $\mu$
problem. Since the MSSM Higgses, $H_u$ and $H_d$, have negative $U(1)_A$ charges, the mass
term is forbidden by the SUSY zero mechanism. With the positively charged singlet field $S$, 
the term $SH_uH_d$ can be allowed, but $\langle S\rangle=0$ in the SUSY limit. However, if
SUSY is broken at the weak scale $\Lambda_W$, the VEV of $S$ becomes non-vanishing and the
order of the weak scale. This results in the SUSY Higgs mass with ${\cal O}(\Lambda_W)$. 
The $b$ parameter also becomes the weak scale.

\subsection{Spontaneous CP violation and discrete symmetry}

As we have seen, one of the attractive features of $E_6\times SU(2)_F$ SUSY GUT is the modified universal sfermion mass (MUSM) (\ref{eq:MUSM}), which evades FCNC constraints with keeping naturalness.
However, such type of mass spectrum is generically suffering from the CEDM constraint of the up quark \cite{Hisano:2004tf}.
Fortunately, we can evade this constraint by introducing a spontaneous CP violation (SCPV) \cite{Lee:1973iz, Barr:1988wk}.
By using SCPV, we can realize real up-type Yukawa couplings together with the KM phase.

Let us introduce the $E_6\times SU(2)_F$ singlet field $S\ (s>0)$.
Then we obtain the following superpotential made of flavon fields $F_a, \bar{F}^a$ and $S$:
\begin{equation}
W=\lambda^sS[\sum_{n=0}^{n_f}c_n\lambda^{(f+\bar{f})n}(\bar{F}F)^n].
\end{equation}
Here $c_n$ are real ${\cal O}(1)$ coefficients and $n_f$ is the number where SUSY-zero mechanism truncates the sum.
When $n_f\geq 2$, the F-flatness condition with respect to $S$ leads to complex VEV $\langle \bar{F}F\rangle$ and then CP symmetry can be spontaneously broken.
Using $SU(2)_F$ gauge symmetry and its D-flatness condition, we can take only $\langle F\rangle$ is complex without loss of generality:
\begin{equation}
\langle F_a\rangle\sim \left(\begin{array}{c}0\\ e^{i\rho}\lambda^{-(f+\bar{f})/2}\end{array}\right), \ \langle \bar{F}^a\rangle\sim \left(\begin{array}{c}0\\ \lambda^{-(f+\bar{f})/2}\end{array}\right).
\end{equation}

Unfortunately, the SCPV affects $\mu$ generation in $U(1)_A$ framework and leads to an unwanted outcome \cite{Ishiduki:2009vr}
\begin{equation}
\mathrm{Arg}[\mu b^\ast]={\cal O}(1),
\end{equation}
if $S\bar FF$ exists.
This rephasing invariant phase induces EDMs of quarks and leptons, and give rise to the SUSY CP problem again.
In order to evade such unwanted relation, we introduce an additional discrete symmetry $Z_6$,
which forbids $S\bar FF$. Then we can avoid this old-type SUSY CP problem. Moreover, 
this symmetry results in the real up-type Yukawa matrix, which solves new-type of SUSY CP 
problem, i.e., the diagonalizing unitary matrices become real and therefore the mass insertion
matrices for up-type quark become real, if the MSSM Higgs $H_u$ is included in $\bm{10}_H$. 
This symmetry constrains the model and reduces the number of ${\cal O}(1)$ parameters.
As a byproduct, the up-quark Yukawa coupling $y_u$ and the CKM matrix element $V_{ub}$ become closer to experimental values $y_u\sim\lambda^8$, $V_{ub}\sim\lambda^4$ than the expectation of $E_6$ GUT $y_u\sim\lambda^6$, $V_{ub}\sim\lambda^3$ \cite{Ishiduki:2009vr}.

It is also reported in \cite{Ishiduki:2009vr} that the discrete symmetry makes one of the neutrinos massless.
This result seems to be unrealistic, but we will show that this issue can be solved by introducing some operators, which have not been considered in the previous paper.

\section{Calculation of neutrino mass and MNS matrix}

\subsection{Field contents and Yukawa couplings}

First, we summarize the field content of the model and its representations under the $E_6$, $SU(2)_F$, $U(1)_A$ and $Z_6$ symmetries.

\begin{table}[tbp]
\begin{center}
\begin{tabular}{c|cccccccccc}
 & $\Psi_a$ & $\Psi_3$ & $F_a$ & $\bar{F}^a$ & $H$ & $\bar{H}$ & $C$ & $\bar{C}$ & $C'$ & $\bar{C}'$\\
\hline
$E_6$ & $\bm{27}$ & $\bm{27}$ & $\bm{1}$ & $\bm{1}$ & $\bm{27}$ & $\bm{\bar{27}}$ & $\bm{27}$ & $\bm{\bar{27}}$ & $\bm{27}$ & $\bm{\bar{27}}$\\
$SU(2)_F$ & $\bm{2}$ & $\bm{1}$ & $\bm{2}$ & $\bar{\bm{2}}$ & $\bm{1}$ & $\bm{1}$ & $\bm{1}$ & $\bm{1}$ & $\bm{1}$ & $\bm{1}$\\
$U(1)_A$ & 4 & $\frac{3}{2}$ & $-\frac{3}{2}$ & $-\frac{5}{2}$ & -3 & 1 & -4 & -1 & 7 & 9\\
$Z_6$ & 0 & 0 & 1 & 0 & 0 & 0 & 5 & 3 & 3 & 3\\
\hline
\end{tabular}
\vspace{5mm}\\
\begin{tabular}{c|ccccccccc}
 & $A$ & $A'$ & $Z_0$ & $Z_2$ & $Z_3$ & $Z_4$ & $S$ & $S'$ & $S''$\\
\hline
$E_6$ & $\bm{78}$ & $\bm{78}$ & $\bm{1}$ & $\bm{1}$ & $\bm{1}$ & $\bm{1}$ & $\bm{1}$ & $\bm{1}$ & $\bm{1}$\\
$SU(2)_F$ & $\bm{1}$ & $\bm{1}$ & $\bm{1}$ & $\bm{1}$ & $\bm{1}$ & $\bm{1}$ & $\bm{1}$ & $\bm{1}$ & $\bm{1}$\\
$U(1)_A$ & -1 & 5 & -1 & -3 & -2 & -5 & 9 & 8 & 5\\
$Z_6$ & 3 & 3 & 0 & 2 & 3 & 4 & 0 & 4 & 4\\
\hline
\end{tabular}
\caption{Field contents and charge assignment under $E_6$$\times SU(2)_F\times U(1)_A\times Z_6$.}
\end{center}
\end{table}

We introduce the following fields, which are listed in Table 1.
All matter fields $\Psi$ belong to $\bm{27}$ representation of $E_6$.
We make its first two generations as a doublet $\Psi_a\ (a=1,2)$ and the third generation as a singlet $\Psi_3$ of the $SU(2)_F$, respectively.
The $SU(2)_F$ and CP are simultaneously broken by the VEVs of the flavon fields $F_a$ and $\bar{F}^a$.
All the other fields are singlets under $SU(2)_F$
\footnote{
To cancel the Witten's anomaly, odd number of additional doublets of $SU(2)_F$ are required.
}. 
$H$ is the field whose VEV $\langle \bm{1}_H\rangle\neq 0$ breaks $E_6$ into $SO(10)$, and $\bar{H}$ is introduced to maintain the D-flatness condition.
The VEV of $A$ breaks $SO(10)$ into $SU(3)_C\times SU(2)_L\times SU(2)_R\times U(1)_{B-L}$.
$C$ is the field whose VEV breaks $SU(3)_C\times SU(2)_L\times SU(2)_R\times U(1)_{B-L}$ into $G_{\rm{SM}}$, and $\bar{C}$ is also introduced to maintain the corresponding D-flatness condition.
Basically, the F-flatness conditions of the positively charged fields, $A'$, $C'$, $\bar C'$ 
etc., determine the VEVs of 
the negatively charged fields.
For example, the F-flatness conditions with respect to $A'$ make $A$ acquire DW-type VEVs for the $SO(10)$ adjoint component to solve the DTS problem.
The alignment between the VEVs of $A$ and $C$, $\bar C$ are realized and at the same time, 
the pseudo Nambu-Goldstone modes become heavy \cite{Barr:1997hq}.
In Table 1, $U(1)_A$ charges are assigned so that the DTS and appropriate Yukawa hierarchies are realized.
Also, $Z_6$ charges are determined so that the SUSY CP problem is evaded.

$S$, $S'$ and $S''$ are introduced to realize the SCPV and $\mu$ generation.
The relevant superpotential are
\begin{eqnarray}
W_S & = & \lambda^{s}S+\lambda^{s+3z_2}SZ_2^3+\lambda^{s+3h}SH^3,\\
W_{S'} & = & \lambda^{s'+z_2}S'Z_2+\lambda^{s'+\bar{c}+c}S'\bar{C}C+\lambda^{s'+2\bar{f}+2f}S'(\bar{F}F)^2,\\
W_{S''} & = & \lambda^{s''+z_2}S''Z_2+\lambda^{s''+\bar{c}+c}S''\bar{C}C.
\end{eqnarray}
First, we obtain the VEV $\langle Z_2\rangle$ from the F-flatness condition of $W_S$ with respect to $S$.
Since the one of the three solutions is real, we assume $\langle Z_2\rangle$ is real.
Second, the F-flatness condition of $W_{S''}$ determines the real VEV $\langle\bar{C}C\rangle$\footnote{We must forbid the term $S^{(')}\bar{F}F\bar{C}C$, which lead the complex VEV $\langle\bar{C}C\rangle$.
If we can choose the basis in which $\langle C\rangle$ is real and $\langle \bar{C}\rangle$ has an opposite phase to $\langle F\rangle$, the up-type Yukawa can keep real.
However, it changes the phase of the down-type Yukawa into removable, and the KM phase cannot be realized.}.
Third, the F-flatness condition of $W_{S'}$ leads the complex VEV $\langle\bar{F}F\rangle$, which breaks the $SU(2)_F$ and CP spontaneously.
Note that the operator $H^3$ gives the SM Higgs mass after developing the VEV of $H$
if the SM Higgses are included in ${\bf 10_H}$.
Then, after breaking SUSY, $W_S$ generates $\mu$ and $b$ terms.
Since $H^3$ does not couple to $S'$ and $S''$, so $\mu$ and $b$ are real.

Let us examine the mass matrices of quarks and leptons in this model.
Under the charge assignment of Table 1, the following interactions between matter and Higgs fields are allowed:
\begin{equation}
Y^H: \quad\left(
\begin{array}{ccc}
0 & d\Psi^a(A,Z_3, \bar HH)\Psi_a & 0\\
d\Psi^a(A,Z_3, \bar HH)\Psi_a & c\lambda^{2(\psi_a+\bar{f})}\bar{F}^a\Psi_a\bar{F}^b\Psi_b & b\lambda^{\psi_a+\psi_3+\bar{f}}\bar{F}^a\Psi_a\Psi_3\\
0 & b\lambda^{\psi_a+\psi_3+\bar{f}}\Psi_3\bar{F}^a\Psi_a & a\lambda^{2\psi_3}\Psi_3\Psi_3
\end{array}
\right)\lambda^hH,
\label{eq:YH}
\end{equation}
\begin{equation}
Y^C: \quad\left(
\begin{array}{ccc}
0 & f'\lambda^{2\psi_a+f+\bar{f}}F^a\Psi_a\bar{F}^b\Psi_b & g'\lambda^{\psi_a+\psi_3+f}F^a\Psi_a\Psi_3\\
f'\lambda^{2\psi_a+f+\bar{f}}\bar{F}^a\Psi_aF^b\Psi_b & 0 & 0\\
g'\lambda^{\psi_a+\psi_3+f}\Psi_3F^a\Psi_a & 0 & 0
\end{array}
\right)\lambda^cC.
\label{eq:YC}
\end{equation}
Here we explicitly write in all ${\cal O}(1)$ coefficients, $a, b, c, d, f'$ and $g'$.
Note that all the ${\cal O}(1)$ coefficients are assumed to be real because of the original CP symmetry.

In (\ref{eq:YH}), the structure of (1,2) and (2,1) elements is a little bit complicate, so we should see this point carefully.
$\Psi^a(A,Z_3, \bar HH)\Psi_a$ consist of several types of terms.
Cleally, some additional field(s) for (1,2) and (2,1) elements are needed because 
$\Psi^a\Psi_a=\epsilon^{ab}\Psi_a\Psi_b=0$, where $\epsilon^{12}=-\epsilon^{21}=1, \epsilon^{11}=\epsilon^{22}=0$.
The negatively $U(1)_A$ charged fields which can have non-vanishing VEVs can be a 
candidate for the additional fields.
There are two possibilities.
One is using the adjoint Higgs $A$: $\Psi^aAZ_3\Psi_a$, $\Psi^aA^2\Psi_a$, etc.
Since the VEV of $A$ breaks $SO(10)$ into 
$SU(3)_C\times SU(2)_L\times SU(2)_R\times U(1)_{B-L}$,
the contributions of these terms are different for the different components of $\Psi_a$
under $SU(3)_C\times SU(2)_L\times SU(2)_R\times U(1)_{B-L}$.
Because the VEV of $A$ is proportional to the $B-L$ charge, there is no contribution 
for the field which has vanishing $B-L$ charge. 
The other possibility, which has not been considered in the previous paper, is using $\bar{H}H$: $\Psi^a\bar{H}H\Psi_a$, etc. Since the VEV of $H$ breaks $E_6$ into $SO(10)$, it is
useful to write down the contributions with the representations of $SO(10)$.
 For example, 
$\epsilon^{ab}\bm{10}_{\Psi_a}\bm{1}_{\Psi_b}$ and 
$\epsilon^{ab}\bm{16}_{\Psi_a}\bm{10}_{\Psi_b}$.
Note that the terms
$\epsilon^{ab}\bm{10}_{\Psi_a}\bm{10}_{\Psi_b}$ and
$\epsilon^{ab}\bm{16}_{\Psi_a}\bm{16}_{\Psi_b}$
are vanishing and give no contribution.
Here we use a conventions for the component fields of $SO(10)$ of a field $X$ as
\begin{eqnarray}
X&=&\bm{27}_X=\bm{16}_X+\bm{10}_X+\bm{1}'_X, \\
\bm{16}_X&=& Q_X+U_X+E_X+D_X+L_X+N_X, \\
\bm{10}_X&=& H^C_X+(H_u)_X+\bar{H}^C_X+(H_d)_X.
\end{eqnarray}
Then, we can write down the concrete components of the interactions 
$\Psi^a(A,Z_3, \bar HH)\Psi_a$ as
\[
d\Psi^a\langle(A, Z_3,\bar HH)\rangle \Psi_aH \ni d_5\lambda^5\epsilon^{ab}H^C_{\Psi_a}\bar{H}^C_{\Psi_b}\bm{1}'_H, \ -\frac{1}{2}d_q\lambda^5\epsilon^{ab}Q_{\Psi_a}U_{\Psi_b}(H_u)_H, 
\]
\begin{equation}
-\frac{1}{2}d_q\lambda^5\epsilon^{ab}Q_{\Psi_a}D_{\Psi_b}(H_d)_H, \ -\frac{3}{2}d_l\lambda^5\epsilon^{ab}L_{\Psi_a}E_{\Psi_b}(H_d)_H, \ h\lambda^5\epsilon^{ab}L_{\Psi_a}N_{\Psi_b}(H_u)_H,
\end{equation}
where $d_5$, $d_q$, $d_l$, $h$ are the ${\cal O}(1)$ coefficients that are different from each other generically.
As we will discuss later, the $h$ term is very important for neutrino masses.

\subsection{Massless modes}

Once we write down the Yukawa couplings, we can calculate the massless combinations of  $\bar{\bm{5}}_i$ and $\bar{\bm{5}}_i'$.
We mainly follow the procedure given in \cite{arXiv:1005.1049}.
First, we fix ${\cal O}(1)$ coefficient of two VEVs:
\begin{equation}
\frac{\lambda^c\langle C\rangle}{\lambda^h\langle H\rangle}\equiv x\lambda^{0.5},
\end{equation}
where $x$ is a real ${\cal O}(1)$ coefficient.
Then, after the Higgs and flavon fields acquire VEVs, (\ref{eq:YH}) and (\ref{eq:YC}) induce following mass matrix for $\bm{5}_i$ and $(\bar{\bm{5}}_i', \bar{\bm{5}}_i)$:
\begin{equation}
\left(
\begin{array}{ccc|ccc}
0 & \alpha d_5\lambda^5 & 0 & 0 & fe^{i\rho}\lambda^{5.5} & ge^{i\rho}\lambda^{3.5}\\
-\alpha d_5\lambda^5 & c\lambda^4 & b\lambda^2 & fe^{i\rho}\lambda^{5.5} & 0 & 0\\
0 & b\lambda^2 & a & ge^{i\rho}\lambda^{3.5} & 0 & 0
\end{array}
\right)
\equiv(M_1\ M_2).
\label{eq:5_mass_matrix}
\end{equation}
Here, we reparametrize ${\cal O}(1)$ coefficients as $f\equiv xf'$ and $g\equiv xg'$.
Each power of $\lambda$ is determined by the corresponding $U(1)_A$ charge.
It is important to note that $\alpha=1$ for the colored Higgs components $(H^C, \bar{H}^{\bar{C}})$ of $\bm{5}$ and $\bar{\bm{5}}'$, and $\alpha=0$ for the doublet Higgs components $(H_u, H_d)$ of those, since $(1, 2)$ and $(2, 1)$ elements of (\ref{eq:5_mass_matrix}) originate from the $B-L$ conserving VEV of $A$.

In order to find the massless combinations, let us diagonalize the $3\times 6$ matrix (\ref{eq:5_mass_matrix}) as follows:
\begin{equation}
V^\dagger(M_1\ M_2)
\left(\begin{array}{cc}
U_{10}^H & U_{10}^0\\
U_{16}^H & U_{16}^0
\end{array}\right)=(M_H^{\rm{diag}}\ 0).
\label{eq:diagonalization}
\end{equation}
Here $V$ is a $3\times 3$ unitary matrix and
\begin{equation}
U\equiv\left(\begin{array}{cc}
U_{10}^H & U_{10}^0\\
U_{16}^H & U_{16}^0
\end{array}\right)
\end{equation}
is a $6\times 6$ unitary matrix, and they rotate $\bm{5}_i$ and $(\bar{\bm{5}}'_i, \bar{\bm{5}}_i)$ $(i=1,2,3)$ into their mass eigenstates, respectively.
Our task is finding matrices $U^0_{10}$ and $U^0_{16}$ which are related by
\begin{equation}
M_1U_{10}^0+M_2U_{16}^0=0.
\label{eq:diagonal_condition}
\end{equation}
According to the calculation in Appendix A, they become
\begin{equation}
U_{10}^0=\left(\begin{array}{ccc}
-\frac{a\alpha d_5(bg-af)}{(ac-b^2)^2}\lambda^{2.5}e^{i\rho} & 1 & {\cal O}(\lambda^{5.5})\\
\frac{bg-af}{ac-b^2}\lambda^{1.5}e^{i\rho} & \frac{a\alpha d_5}{ac-b^2}\lambda &{\cal O}(\lambda^{4.5})\\
-(\frac{g}{a}+\frac{b}{a}\frac{bg-af}{ac-b^2})\lambda^{3.5}e^{i\rho} & -\frac{b\alpha d_5}{ac-b^2}\lambda^3 & {\cal O}(\lambda^{6.5})
\end{array}\right),
\label{eq:U10}
\end{equation}
\begin{equation}
U_{16}^0=\left(\begin{array}{ccc}
1 & 0 & 0\\
{\cal O}(\lambda^6) & 0 & 1\\
-\frac{bg-af}{ac-b^2}\frac{\alpha d_5}{g}\lambda^3 & -\frac{\alpha d_5^2}{ac-b^2}\frac{a}{g}\lambda^{2.5}e^{-i\rho} & -\frac{f}{g}\lambda^2
\end{array}\right).
\label{eq:U16}
\end{equation}
The calculation is taken at the leading order, but $(U^0_{10})_{13, 23, 33}$ and $(U^0_{16})_{21}$ are calculated at the next-leading order because the corresponding leading terms are cancelled.
Therefore, the massless modes of $\bar{\bm{5}}$ can be written as
\begin{equation}
\bar{\bm{5}}^0_i\equiv(U_{10}^{0\dagger})_{ij}\bar{\bm{5}}'_j+(U_{16}^{0\dagger})_{ij}\bar{\bm{5}}_j
=\left(\begin{array}{c}
\bar{\bm{5}}_1+\cdots\\ \bar{\bm{5}}'_1+\cdots\\ \bar{\bm{5}}_2+\cdots
\end{array}\right).
\label{eq:massless}
\end{equation}

In order to calculate neutrino mass, we should also specify the MSSM Higgs doublets $H_u$ and $H_d$.
They are the combinations which remain massless at the GUT scale.
According to the Appendix A in \cite{Ishiduki:2009vr}, they are
\begin{equation}
H_u\subset \bm{5}_H
\end{equation}
\begin{equation}
H_d\subset \bar{\bm{5}}'_H+\beta_He^{-i\rho}\lambda^{0.5}\bar{\bm{5}}_C.
\label{eq:HD}
\end{equation}
Here $\beta_H$ is a real ${\cal O}(1)$ coefficient and the phase $e^{-i\rho}$ comes from VEVs of the $F$ and $Z_4$\footnote{If we choose $z_2=-1$ (or $z_2=-2$), then the term $\bar{C}'Z_2^5C$ (or $\bar{C}'Z_2^2C$) changes the phase in (\ref{eq:HD}) into arbitrary.
For detailed discussion of the Higgs sector, see the Appendix A in \cite{Ishiduki:2009vr}.}.
As mentioned in \cite{Ishiduki:2009vr}, the second term in (\ref{eq:HD}) is important to ensure the KM phase and make all charged leptons massive.

\subsection{Neutrino mass}

Now we calculate neutrino masses in our model.
In this subsection we explicitly write the cutoff scale $\Lambda$.
First, we compute the Dirac neutrino mass matrix. Since we have 6 right-handed neutrino,
$\bm{1}_i$ and $\bm{1}'_i$, from the interactions ${\bf 10}_i{\bf 1}'_j{\bf 10}_H$
and ${\bf 16}_i{\bf 16}_j{\bf 10}_H$, 
$3\times 6$ Dirac neutrino Yukawa matrix $Y_{\nu_D}$ is given by
\begin{equation}
Y_{\nu_D}= (Y_{\nu_D1'}\ Y_{\nu_D1})=(U_{10}^{0T}(Y^H_{1'})^T\ U_{16}^{0T}(Y^H_1)^T),
\label{eq:Dirac_mass}
\end{equation}
where
\begin{equation}
Y^H_{1'}=\left(\begin{array}{ccc}
0 & h\lambda^5 & 0\\
-h\lambda^5 & c\lambda^4 & b\lambda^2\\
0 & b\lambda^2 & a
\end{array}\right), \quad
Y^H_1=\left(\begin{array}{ccc}
0 & -\frac{3}{2}d_l\lambda^5 & 0\\
\frac{3}{2}d_l\lambda^5 & c\lambda^4 & b\lambda^2\\
0 & b\lambda^2 & a
\end{array}\right).
\end{equation}
It is also important that $\alpha=0$ in the mixing matrices (\ref{eq:U10}) (\ref{eq:U16}).
Using these facts, we can express the Dirac mass as follows:
\begin{equation}
Y_{\nu_D}=\left(\begin{array}{ccc|ccc}
h\frac{bg-af}{ac-b^2}\lambda^{6.5}e^{i\rho} & -f\lambda^{5.5}e^{i\rho} & -g\lambda^{3.5}e^{i\rho} & 0 & -\frac{3}{2}d_l\lambda^5 & 0\\
0 & -h\lambda^5 & 0 & 0 & 0 & 0\\
0 & 0 & 0 & \frac{3}{2}d_l\lambda^5 & (c-\frac{bf}{g})\lambda^4 & \frac{bg-af}{g}\lambda^2
\end{array}\right).
\end{equation}
Note that the second generation would be massless without $h$.
The contribution from the term $\Psi^a\bar{H}H\Psi_aH$ is essential for obtaining the realistic neutrino sector.

Second, we introduce the $6\times 6$ Majorana right-handed neutrino mass matrix.
Since each $\bm{27}$ has two $\bm{1}$'s, the Majorana mass has following three types of contribution:
\begin{eqnarray}
W & = & \frac{(Y_{1'1'})_{ij}}{\Lambda}\Psi_i\Psi_j\bar{H}\bar{H}+\frac{(Y_{1'1})_{ij}}{\Lambda}\Psi_i\Psi_j\bar{H}\bar{C}+\frac{(Y_{11})_{ij}}{\Lambda}\Psi_i\Psi_j\bar{C}\bar{C}\nonumber\\
 & \to &  \frac{(Y_{1'1'})_{ij}}{\Lambda}\bm{1}'_i\bm{1}'_j\langle\bm{1}'_{\bar{H}}\rangle\langle\bm{1}'_{\bar{H}}\rangle+\frac{(Y_{1'1})_{ij}}{\Lambda}\bm{1}'_i\bm{1}_j\langle\bm{1}'_{\bar{H}}\rangle\langle\bm{1}_{\bar{C}}\rangle+\frac{(Y_{11})_{ij}}{\Lambda}\bm{1}_i\bm{1}_j\langle\bm{1}_{\bar{C}}\rangle\langle\bm{1}_{\bar{C}}\rangle.
 \end{eqnarray}
Then the Majorana mass term is $6\times 6$ matrix
\begin{equation}
(\bm{1}'_i\ \bm{1}_i)\left(Y_{N_R^c}\right)_{ij}\left(\begin{array}{c}\bm{1}'_j\\ \bm{1}_j\end{array}\right)\frac{\langle\bar{H}\rangle^2}{\Lambda},\qquad
Y_{N_R}=\left(\begin{array}{cc}
Y_{1'1'} & Y_{1'1}\\
Y_{1'1}^T & Y_{11}
\end{array}\right).
\end{equation}
Here, we parametrize $Y_{N_R}^{-1}$ with ${\cal O}(1)$ coefficients as
\begin{equation}
Y_{N_R}^{-1}=\left(\begin{array}{ccc|ccc}
N_{11}\lambda^{-13} & N_{12}\lambda^{-12} & N_{13}\lambda^{-10} & N_{14}\lambda^{-12.5} & N_{15}\lambda^{-11.5} & N_{16}\lambda^{-9.5}\\
N_{12}\lambda^{-12} & N_{22}\lambda^{-11} & N_{23}\lambda^{-9} & N_{24}\lambda^{-11.5} & N_{25}\lambda^{-10.5} & N_{26}\lambda^{-8.5}\\
N_{13}\lambda^{-10} & N_{23}\lambda^{-9} & N_{33}\lambda^{-7} & N_{34}\lambda^{-9.5} & N_{35}\lambda^{-8.5} & N_{36}\lambda^{-6.5}\\
\hline
N_{14}\lambda^{-12.5} & N_{24}\lambda^{-11.5} & N_{34}\lambda^{-9.5} & N_{44}\lambda^{-12} & N_{45}\lambda^{-11} & N_{46}\lambda^{-9}\\
N_{15}\lambda^{-11.5} & N_{25}\lambda^{-10.5} & N_{35}\lambda^{-8.5} & N_{45}\lambda^{-11} & N_{55}\lambda^{-10} & N_{56}\lambda^{-8}\\
N_{16}\lambda^{-9.5} & N_{26}\lambda^{-8.5} & N_{36}\lambda^{-6.5} & N_{46}\lambda^{-9} & N_{56}\lambda^{-8} & N_{66}\lambda^{-6}\\
\end{array}\right),
\end{equation}
where $N_{ij}$ are ${\cal O}(1)$ complex parameters.

Finally we can calculate the neutrino mass using the seesaw mechanism.
The result is
\begin{equation}
M_\nu\equiv Y_\nu\frac{\langle H_u\rangle^2}{\Lambda}, \qquad Y_\nu  =  Y_{\nu_D}Y_{N_R}^{-1}Y_{\nu_D}^T=\left(\begin{array}{ccc}
y_{11}\lambda^{-1} & y_{12}\lambda^{-1.5} & y_{13}\lambda^{-2}\\
y_{12}\lambda^{-1.5} & y_{22}\lambda^{-2} & y_{23}\lambda^{-2.5}\\
y_{13}\lambda^{-2} & y_{23}\lambda^{-2.5} & y_{33}\lambda^{-3}\\
\end{array}\right),
\end{equation}
where $y_{ij}$ are complicate combinations of ${\cal O}(1)$ coefficient: 
\[
y_{11}=N_{11}(h\frac{bg-af}{ac-b^2})^2e^{2i\rho}+N_{22}f^2e^{2i\rho}+N_{33}g^2e^{2i\rho}+N_{55}(-\frac{3}{2}d_l)^2
\]
\[
\qquad\qquad -2N_{12}fh\frac{bg-af}{ac-b^2}e^{2i\rho}-2N_{13}gh\frac{bg-af}{ac-b^2}e^{2i\rho}-2N_{15}h(-\frac{3}{2}d_l)\frac{bg-af}{ac-b^2}
\]
\begin{equation}
\qquad\qquad +2N_{23}fge^{2i\rho}-2N_{25}f(-\frac{3}{2}d_l)e^{i\rho}-2N_{35}g(-\frac{3}{2}d_l)e^{i\rho},
\end{equation}
\begin{equation}
y_{12}=-h\left[N_{12}h\frac{bg-af}{ac-b^2}e^{i\rho}-N_{22}fe^{i\rho}-N_{23}ge^{i\rho}+N_{25}(-\frac{3}{2}d_l)\right],
\end{equation}
\[
y_{13}=h\frac{bg-af}{ac-b^2}e^{i\rho}\left[N_{14}(\frac{3}{2}d_l)+N_{15}(\frac{ac-b^2}{a}+\frac{bg-af}{b}\frac{b}{a})+N_{16}\frac{bg-af}{g}\right]
\]
\[
-fe^{i\rho}\left[N_{24}(\frac{3}{2}d_l)+N_{25}(\frac{ac-b^2}{a}+\frac{bg-af}{b}\frac{b}{a})+N_{26}\frac{bg-af}{g}\right]
\]
\[
-ge^{i\rho}\left[N_{34}(\frac{3}{2}d_l)+N_{35}(\frac{ac-b^2}{a}+\frac{bg-af}{b}\frac{b}{a})+N_{36}\frac{bg-af}{g}\right]
\]
\begin{equation}
-\frac{3}{2}d_l\left[N_{45}(\frac{3}{2}d_l)+N_{55}(\frac{ac-b^2}{a}+\frac{bg-af}{b}\frac{b}{a})+N_{56}\frac{bg-af}{g}\right],
\end{equation}
\begin{equation}
y_{22}=h^2N_{22},
\end{equation}
\begin{equation}
y_{23}=-h\left[N_{24}(\frac{3}{2}d_l)+N_{25}(\frac{ac-b^2}{a}+\frac{bg-af}{b}\frac{b}{a})+N_{26}\frac{bg-af}{g}\right],
\end{equation}
\[
y_{33}=N_{44}(\frac{3}{2}d_l)^2+N_{55}(\frac{ac-b^2}{a}+\frac{bg-af}{b}\frac{b}{a})^2+2N_{46}(\frac{3}{2}d_l)\frac{bg-af}{g}
\]
\begin{equation}
+2N_{56}\frac{bg-af}{g}(\frac{ac-b^2}{a}+\frac{bg-af}{b}\frac{b}{a}).
\end{equation}

These masses give $\Delta m_{12}^2/\Delta m_{23}^2\sim \lambda^2=4.8\times 10^{-2}$, which
is consistent with the experimental facts $\Delta m_{12}^2/\Delta m_{23}^2\simeq 3.1\times 10^{-2}$ \cite{PDG} up to the combination of ${\cal O}(1)$ coefficients.

\subsection{MNS matrix}

We can also calculate the MNS matrix in our model.
First, charged lepton Yukawa matrix is given \cite{Ishiduki:2009vr} as
\begin{equation}
Y_e=\left(\begin{array}{ccc}
[\frac{bg-af}{ac-b^2}(f'+\frac{bg'}{a})-\frac{gg'}{a}]\beta_He^{2i\delta}\lambda^6 & \frac{3}{2}d_l\lambda^5 & 0\\
0 & \beta_Hf'e^{i\delta}\lambda^{4.5} & \beta_Hg'e^{i\delta}\lambda^{2.5}\\
-\frac{3}{2}d_l\lambda^5 & (\frac{ac-b^2}{a}+\frac{bg-af}{g}\frac{b}{a})\lambda^4 & \frac{bg-af}{g}\lambda^2
\end{array}\right).
\end{equation}
Then we can calculate MNS matrix.
Following the diagonal procedure given in \cite{King:2002nf}, it is computed, at the leading order, 
\begin{equation}
V_{\rm{MNS}}=\left(\begin{array}{ccc}
1 & v_{12}\lambda^{0.5} & v_{13}\lambda\\
-v_{12}^\ast\lambda^{0.5} & 1 & v_{23}\lambda^{0.5}\\
(v_{12}^\ast v_{23}^\ast -v_{13}^\ast)\lambda & -v_{23}^\ast\lambda^{0.5} & 1
\end{array}\right),
\end{equation}
where $v_{12}, v_{13}$, and $v_{23}$ are written as
\begin{equation}
v_{12}=\frac{\frac{3}{2}d_l}{\beta_H}\frac{1}{1-gg'\frac{bf+c}{bg-af}}-\frac{y_{12}y_{33}-y_{13}y_{23}}{y_{22}y_{33}-y_{23}^2},
\end{equation}
\begin{equation}
v_{13}=-\frac{y_{13}}{y_{33}}-v_{23}^\ast\frac{y_{12}y_{33}-y_{13}y_{23}}{y_{22}y_{33}-y_{23}^2},
\end{equation}
\begin{equation}
v_{23}=\frac{\beta_Hgg'}{bg-af}-\frac{y_{23}}{y_{33}}.
\end{equation}

This result can be compared with experimental values.
The model says $\tan\theta_{12}=V_{12}/V_{11}\sim \lambda^{0.5}=0.47$, $\tan\theta_{23}=V_{23}/V_{33}\sim \lambda^{0.5}=0.47$ up to the combination of ${\cal O}(1)$ coefficients.
These are consistent with experiment $\tan\theta_{12}\simeq 0.68$, $\tan\theta_{23}\geq 0.747$ \cite{PDG}.
It is also interesting that recent T2K result \cite{Abe:2011sj} $0.087<\sin\theta_{13}<0.275$ (assuming $\delta_{CP}=0$) supports our result $\sin\theta_{13}=|V_{13}|\sim \lambda=0.22$, up to the combination of ${\cal O}(1)$ coefficients.

\subsection{The role of the $\Psi\bar{H}H\Psi H$ terms}

As we have seen, the terms $\Psi^a\bar{H}H\Psi_a H$ played the important role for the 
non-vanishing second generation neutrino mass.
Here we show that this type of higher-dimensional operators do not affect other structures of this model, for example, the quark masses and charged lepton masses.

First, we classify $SO(10)$ invariant Yukawa interaction derived from $E_6$ invariant Yukawa term $\Psi_i\Psi_jH$ (or $\Psi_i\Psi_jC$):
\begin{equation}
\bm{27}_i\bm{27}_j\bm{27}_{H,C}=\bm{16}_i\bm{16}_j\bm{10}_{H,C}+\bm{10}_i\bm{10}_j\bm{1}_{H,C}+\bm{16}_i\bm{10}_j\bm{16}_{H,C}+\bm{10}_i\bm{1}_j\bm{10}_{H,C}.
\label{eq:E6toSO(10)}
\end{equation}
Note that these terms have common ${\cal O}(1)$ coefficients, and all terms are symmetric between $i$ and $j$, although at $SO(10)$ level, the coefficients of the terms 
$\bm{16}_i\bm{10}_j\bm{16}_{H,C}$ and $\bm{10}_i\bm{1}_j\bm{10}_{H,C}$ do not have to be taken as symmetric.

This property changes if we include higher-dimensional operators like $\Psi_i\bar{H}H\Psi_jH$ or $\Psi_i\bar{H}H\Psi_jC$.
Each terms in the right-hand-side of (\ref{eq:E6toSO(10)}) receives extra contribution, which generically differs from each other.
Therefore, they cannot have common ${\cal O}(1)$ coefficients, and the number of parameters increases.

Fortunately, in our model, only (1,2) and (2,1) elements of $Y^H$ have room of $U(1)_A$ charge for including $\bar{H}H$, and in other Yukawa elements, SUSY-zero forbids such higher-dimensional operators.
Since these elements are anti-symmetric due to $SU(2)_F$, the extra contributions $\bm{16}_i\bm{16}_j\bm{10}_H$ and $\bm{10}_i\bm{10}_j\bm{1}_H$ vanish.
Therefore, the higher-dimensional operator including $\bar{H}H$ only affects the Dirac neutrino mass through $\bm{10}_i\bm{1}_j\bm{10}_H$, while the quark sector and charged lepton sector is unchanged.

According to the previous paper \cite{Ishiduki:2009vr}, the model has characteristic predictions $V_{ub}\sim \lambda^4$ and $V_{cb}y_b=y_c$ at the GUT scale.
We will show in Appendix B that these predictions are not spoiled even if the model generically contains such higher-dimensional operators. Sufficient conditions for these 
predictions are $\langle A\rangle\propto Q_{B-L}$ and that the (2,2), (2,3), (3,2), and (3,3)
components of $Y^C$ are vanishing. Such conditions are satisfied in more general $E_6$ GUTs
with family symmetry. The former condition should be satisfied to solve the doublet-triplet
splitting problem, and the latter can be satisfied by the SUSY zero mechanism if the 
anomalous $U(1)_A$ charge of $C$ is smaller than that of $H$. 
The discrete symmetry for solving the SUSY CP problem is not necessary.
Therefore, these two predictions are rather general ones. The first prediction
$V_{ub}\sim \lambda^4$ has been tested by B-factory experiments, and the second
prediction $V_{cb}y_b=y_c$ can be tested in future experiments by measuring
the $\tan\beta\equiv \langle H_u\rangle/\langle H_d\rangle $.

\section{Summary and discussion}

In this paper we conducted a detail calculation of the neutrino sector in $E_6\times SU(2)_F$ SUSY GUT with spontaneous CP violation.
Originally, $E_6$ GUT can explain the realistic neutrino sector as well as quark and charged lepton sector.
However, the previous paper reported the problem about the neutrino mass because of the new discrete symmetry, which is introduced in order to solve the SUSY CP problem.
We computed the neutrino masses and the MNS matrix explicitly and found that the term like $\Psi_i\bar{H}H\Psi_jH$, which have not been considered in the previous paper, is important to make the second generation massive.
These higher dimensional operators only contribute in the neutrino sector, so the structures of quarks and charged leptons, which calculated in the previous paper, are kept valid.

Our result reproduces experimental value well.
$\Delta m_{12}^2/\Delta m_{23}^2$, $\tan\theta_{12}$, $\tan\theta_{23}$ and $\sin\theta_{13}$ are all consistent with experimental values up to the combination of ${\cal O}(1)$ coefficients.
Combining the previous paper and our result, the concrete model reproducing realistic masses and mixings for all the SM fermions is constructed. Note that even if the model has the 
``modified universal sfermion mass spectrum", we have no SUSY CP problem.

In this model, because of the strong constraint of Yukawa structure, there are characteristic predictions $V_{ub}\sim \lambda^4$ and $V_{cb}y_b=y_c$ at the GUT scale \cite{Ishiduki:2009vr}.
These predictions are valid even if we generically include the higher-dimensional operators $\Psi_i\bar{H}H\Psi_jH$ and $\Psi_i\bar{H}H\Psi_jC$.
Unfortunately, we could  get no characteristic prediction in the neutrino sector because of huge number of parameters in the Majorana neutrino mass terms.

\section*{Acknowledgments}
N.M. is supported in part by Grants-in-Aid for Scientific Research from
MEXT of Japan.
This work was partially supported by the Grand-in-Aid for Nagoya
University Global COE Program,
``Quest for Fundamental Principles in the Universe:
from Particles to the Solar System and the Cosmos'',
from the MEXT of Japan.

\appendix

\section{Diagonalization procedure of the superheavy Yukawa matrix}

In this appendix, we derive the expression for the two matrices $U^0_{10}$ (\ref{eq:U10}) and $U^0_{16}$ (\ref{eq:U16}) which describe the massless $\bar{\bm{5}}^0_i$ as the combination of $\bar{\bm{5}}_i$ and $\bar{\bm{5}}'_i$ as (\ref{eq:massless}).
All calculation is performed at the leading order.

First, we show the strategy of finding $U$ which diagonalize the superheavy Yukawa matrix as (\ref{eq:diagonalization}). In the following, we assume that $M_1$ has rank 3, and therefore
we have a inverse matrix $M_1^{-1}$. 
Then we calculate $M_1^{-1}M_2$ and parametrize it as
\begin{equation}
M_1^{-1}M_2\equiv C\equiv \left(\begin{array}{ccc}
c_{11}\lambda^{0.5} & c_{12}\lambda^{-0.5} & c_{13}\lambda^{-2.5}\\
c_{21}\lambda^{1.5} & c_{22}\lambda^{0.5} & c_{23}\lambda^{-1.5}\\
c_{31}\lambda^{3.5} & c_{32}\lambda^{2.5} & c_{33}\lambda^{0.5}
\end{array}\right).
\end{equation}
Here $c_{ij}$ are the combination of ${\cal O}(1)$ coefficients.
After expressing $U^0_{10}$ and $U^0_{16}$ in terms of $c_{ij}$, we will substitute concrete parameters in them.
Then we rotate $M_1^{-1}(M_1\ M_2)=(1_{3\times 3}\ C)$ as
\begin{equation}
(1_{3\times 3}\ C)U'=(C'\ 0_{3\times 3}),
\end{equation}
where $U'$ is a $6\times 6$ unitary matrix
\begin{equation}
U'\equiv\left(\begin{array}{cc}
U_1 & U_3\\
U_2 & U_4
\end{array}\right)
\end{equation}
and $C'$ is a $3\times 3$ matrix.
If we find unitary matrices $V$ and $V'$ which diagonalize $M_1C'$ as
\begin{equation}
V^\dagger M_1C' V'=M_H^{\rm{diag}},
\end{equation}
then we can express the matrix $U$ as
\begin{equation}
U\equiv \left(\begin{array}{cc}
U_{10}^H & U_{10}^0\\
U_{16}^H & U_{16}^0
\end{array}\right)
=U'\left(\begin{array}{cc}
V' & 0\\
0 & T
\end{array}\right)
=\left(\begin{array}{cc}
U_1V' & U_3T\\
U_2V' & U_4T
\end{array}\right).
\end{equation}
Here $T$ is a unitary matrix which corresponds to the degree of freedom of the rotation for the three 0's with keeping (\ref{eq:diagonalization}).

Since we are only interested in $U^0_{10}$ and $U^0_{16}$ in this paper, our task is to find the matrix $U'$.
After tedious calculation, we can find $C'$ and $U'$ as
\begin{equation}
(1_{3\times 3}\ C)U'=\left(\begin{array}{ccc|ccc}
c_{31}\lambda^{-2.5} & 0 & 0 & 0 & 0 & 0\\
c_{32}\lambda^{-1.5} & 1 & 0 & 0 & 0 & 0\\
c_{33}\lambda^{0.5} & (c_{22}-\frac{c_{12}c_{23}}{c_{13}})(c_{32}-\frac{c_{12}c_{33}}{c_{13}})\lambda^{3} & 1 & 0 & 0 & 0
\end{array}\right),
\end{equation}
\[
U'=\left(\begin{array}{ccc|}
\frac{1}{c_{13}}\lambda^{2.5} & -\frac{c_{23}}{c_{13}}\lambda & -\frac{c_{33}}{c_{13}}\lambda^3\\
0 & 1 & -(c_{22}-\frac{c_{12}c_{23}}{c_{13}})(c_{32}-)\lambda^3\\
0 & 0 & 1\\
\hline
\frac{c_{11}}{c_{13}}\lambda^3 & (c_{21}-\frac{c_{11}c_{23}}{c_{13}})\lambda^{1.5} & (c_{31}-\frac{c_{11}c_{33}}{c_{13}})\lambda^{3.5}\\
\frac{c_{12}}{c_{13}}\lambda^2 & (c_{22}-\frac{c_{12}c_{23}}{c_{13}})\lambda^{0.5} & (c_{32}-\frac{c_{12}c_{33}}{c_{13}})\lambda^{2.5}\\
1 & -\frac{c_{12}}{c_{13}}(c_{22}-\frac{c_{12}c_{23}}{c_{13}})\lambda^{2.5} & -\frac{c_{12}}{c_{13}}(c_{32}-\frac{c_{12}c_{33}}{c_{13}})\lambda^{4.5}
\end{array}\right.
\]
\begin{equation}
\left.\begin{array}{|ccc}
\frac{c_{23}}{c_{13}}(c_{21}-\frac{c_{11}c_{23}}{c_{13}})\lambda^{2.5} & 1 & 0\\
-(c_{21}-\frac{c_{11}c_{23}}{c_{13}})\lambda^{1.5} & \frac{c_{23}}{c_{13}}\lambda & -(c_{22}-\frac{c_{12}c_{23}}{c_{13}})\lambda^{0.5}\\
-(c_{31}-\frac{c_{11}c_{33}}{c_{13}})\lambda^{3.5} & \frac{c_{33}}{c_{13}}\lambda^3 & -(c_{32}-\frac{c_{12}c_{33}}{c_{13}})\lambda^{2.5}\\
\hline
1 & 0 & 0\\
-(c_{21}-\frac{c_{11}c_{23}}{c_{13}})(c_{22}-\frac{c_{12}c_{23}}{c_{13}})\lambda^2 & \frac{c_{23}}{c_{13}}(c_{22}-\frac{c_{12}c_{23}}{c_{13}})\lambda^{1.5} & 1\\
-\frac{c_{11}}{c_{13}}\lambda^3 & -\frac{1}{c_{13}}\lambda^{2.5} & -\frac{c_{12}}{c_{13}}\lambda^2
\end{array}\right).
\end{equation}
We choose a basis which $(U^0_{16})_{12}$,  $(U^0_{16})_{13}$ and  $(U^0_{16})_{22}$ are zero by the rotation
\begin{equation}
T=\left(\begin{array}{ccc}
1 & 0 & 0\\
0 & 1 & \frac{c_{23}}{c_{13}}(c_{22}-\frac{c_{12}c_{23}}{c_{13}})\lambda^{1.5}\\
0 & -\frac{c_{23}}{c_{13}}(c_{22}-\frac{c_{12}c_{23}}{c_{13}})\lambda^{1.5} & 1
\end{array}\right).
\end{equation}
After that, we can get $U^0_{10}$ and $U^0_{16}$ in terms of $c_{ij}$ as
\begin{equation}
U_{10}^0=\left(\begin{array}{ccc}
\frac{c_{23}}{c_{13}}(c_{21}-\frac{c_{11}c_{23}}{c_{13}})\lambda^{2.5} & 1 & \frac{c_{23}}{c_{13}}(c_{22}-\frac{c_{12}c_{23}}{c_{13}})\lambda^{1.5}\\
-(c_{21}-\frac{c_{11}c_{23}}{c_{13}})\lambda^{1.5} & \frac{c_{23}}{c_{13}}\lambda & -(c_{22}-\frac{c_{12}c_{23}}{c_{13}})\lambda^{0.5}\\
-(c_{31}-\frac{c_{11}c_{33}}{c_{13}})\lambda^{3.5} & -\frac{c_{33}}{c_{13}}\lambda^{3} & (c_{32}-\frac{c_{12}c_{33}}{c_{13}})\lambda^{2.5}
\end{array}\right),
\end{equation}
\begin{equation}
U_{16}^0=\left(\begin{array}{ccc}
1 & 0 & 0\\
-(c_{21}-\frac{c_{11}c_{23}}{c_{13}})(c_{22}-\frac{c_{12}c_{23}}{c_{13}})\lambda^2 & 0 & 1\\
-\frac{c_{11}}{c_{13}}\lambda^{3} & -\frac{1}{c_{13}}\lambda^{2.5} & -\frac{c_{12}}{c_{13}}\lambda^{2}
\end{array}\right).
\end{equation}
We can check these matrices satisfy the relation (\ref{eq:diagonal_condition}) at the leading order.
Finally, we obtain (\ref{eq:U10}) and (\ref{eq:U16}) by substituting the concrete expressions in $c_{ij}$:
\begin{equation}
c_{11}=\frac{bg-af}{a\alpha d_5}, \ c_{12}=\frac{ac-b^2}{\alpha d_5^2}\frac{f}{a}, \ c_{13}=\frac{ac-b^2}{\alpha d_5^2}\frac{g}{a}
\end{equation}
\begin{equation}
c_{21}=0, \ c_{22}=\frac{f}{\alpha d_5}, \ c_{23}=\frac{g}{\alpha d_5}
\end{equation}
\begin{equation}
c_{31}=\frac{g}{a}, \ c_{32}=-\frac{bf}{a\alpha d_5}, \ c_{33}=-\frac{bg}{a\alpha d_5}
\end{equation}
\section{The condition of the predictions}

In section 3, we obtained the realistic neutrino sector by introducing the higher-dimensional operator $\Psi^a\bar{H}H\Psi_aH$, and saw that such operators cannot be written in other
 elements of $Y^H$ and $Y^C$ because of the SUSY zero mechanism.
In this appendix, we will clarify the conditions of the characteristic predictions $V_{ub}\sim \lambda^4$ and $V_{cb}y_b=y_c$, and show that the conditions can be satisfied even if we generically include the higher-dimensional operators $\Psi_i\bar{H}H\Psi_jH$ and 
$\Psi_i\bar{H}H\Psi_jC$.

In order to see the essence of the predictions, let us choose the basis where the up-type Yukawa matrix $Y_u$ is diagonal.
In this basis, $V_{ub}$ is simply written only by the down-type Yukawa matrix as $V_{ub}\sim (Y_d)_{13}/(Y_d)_{33}$.
The relation $V_{cb}y_b=y_c$ is also simplified as $(Y_d)_{23}=(Y_u)_{22}$.
Therefore, we focus on the quantities $(Y_d)_{13}$ and $(Y_d)_{23}$.

Let us determine the third generation of massless $\bar{\bm{5}}$ combination $\bar{\bm{5}}^0_3$, which is mainly $\bar{\bm{5}}_2$, in this basis.
Since we can choose the basis where $\bar{\bm{5}}^0_3$ does not contain $\bar{\bm{5}}_1$ and $\bar{\bm{5}}'_1$, we parametrize
\begin{equation}
\bar{\bm{5}}^0_3\equiv\frac{1}{\sqrt{1+|\alpha|^2+|\beta|^2+|\gamma|^2}}(\bar{\bm{5}}_2+\alpha\bar{\bm{5}}'_2+\beta\bar{\bm{5}}'_3+\gamma\bar{\bm{5}}_3).
\end{equation}
As we saw in Section 3.2, the massless modes are determined by diagonalizing the $3\times 6$ mass matrix $(M_1\ M_2)$.
Both $Y_u$ and $M_1$ is obtained from $Y^H$, but here we consider general case where these two matrices cannot be diagonalized simultaneously.
In this case, $M_1$ is the form
\begin{equation}
M_1\sim
\left(\begin{array}{ccc}
1 & \lambda & \lambda^3\\
\lambda & 1 & \lambda^2\\
\lambda^3 & \lambda^2 & 1
\end{array}\right)
\left(\begin{array}{ccc}
\lambda^6 & \lambda^5 & \lambda^3\\
\lambda^5 & \lambda^4 & \lambda^2\\
\lambda^3 & \lambda^2 & 1
\end{array}\right)
\left(\begin{array}{ccc}
1 & \lambda & \lambda^3\\
\lambda & 1 & \lambda^2\\
\lambda^3 & \lambda^2 & 1
\end{array}\right)
\sim
\left(\begin{array}{ccc}
\lambda^6 & \lambda^5 & \lambda^3\\
\lambda^5 & \lambda^4 & \lambda^2\\
\lambda^3 & \lambda^2 & 1
\end{array}\right).
\end{equation}
On the other hand, $M_2$ is obtained from $Y^C$.
Again, we consider general case
\begin{equation}
M_2\sim \lambda^{0.5}
\left(\begin{array}{ccc}
1 & \lambda & \lambda^3\\
\lambda & 1 & \lambda^2\\
\lambda^3 & \lambda^2 & 1
\end{array}\right)
\left(\begin{array}{ccc}
\lambda^6 & \lambda^5 & \lambda^3\\
\lambda^5 & \lambda^4 & \lambda^2\\
\lambda^3 & \lambda^2 & 1
\end{array}\right)
\left(\begin{array}{ccc}
1 & \lambda & \lambda^3\\
\lambda & 1 & \lambda^2\\
\lambda^3 & \lambda^2 & 1
\end{array}\right)
\sim
\left(\begin{array}{ccc}
\lambda^{6.5} & \lambda^{5.5} & \lambda^{3.5}\\
\lambda^{5.5} & \lambda^{4.5} & \lambda^{2.5}\\
\lambda^{3.5} & \lambda^{2.5} & \lambda^{0.5}
\end{array}\right).
\end{equation}
Then we can obtain the massless combination of $\bar{\bm{5}}^0_3$ by solving
\begin{equation}
(M_1\ M_2)\left(\begin{array}{c}0\\ \alpha\\ \beta\\ 0\\ 1\\ \gamma\end{array}\right)=0.
\end{equation}
The solution becomes generically
\begin{equation}
\alpha\sim\lambda^{0.5}, \ \beta\sim\lambda^{2,5}, \ \gamma\sim\lambda^2.
\label{eq:5bar_solution}
\end{equation}

Let us evaluate $(Y_d)_{13}$ and $(Y_d)_{23}$ to obtain the conditions for 
$V_{ub}\sim \lambda^4$ and $V_{cb}y_b=y_c$.
The down-type Yukawa matrix $Y_d$ consists of two types of interaction: one is the interaction between $\bm{10}$ and $\bar{\bm{5}}$ via $Y^H_{\rm{down}}$, the other is between $\bm{10}$ and $\bar{\bm{5}}'$ via $\lambda^{0.5}Y^C$. Here we denote the matrix $Y^H$ appearing in the 
down-type quark Yukawa matrix as $Y^H_{\rm{down}}$.
$Y^C$ is written in this basis as
\begin{equation}
Y^C\sim
\left(\begin{array}{ccc}
1 & \lambda & \lambda^3\\
\lambda & 1 & \lambda^2\\
\lambda^3 & \lambda^2 & 1
\end{array}\right)
\left(\begin{array}{ccc}
\lambda^5 & \lambda^4 & \lambda^2\\
\lambda^4 & \lambda^3 & \lambda\\
\lambda^2 & \lambda & \lambda^{-1}
\end{array}\right)
\left(\begin{array}{ccc}
1 & \lambda & \lambda^3\\
\lambda & 1 & \lambda^2\\
\lambda^3 & \lambda^2 & 1
\end{array}\right)
\sim
\left(\begin{array}{ccc}
\lambda^{5} & \lambda^{4} & \lambda^{2}\\
\lambda^{4} & \lambda^{3} & \lambda\\
\lambda^{2} & \lambda & \lambda^{-1}
\end{array}\right).
\end{equation}
For example, the leading contributions for 
$(Y_d)_{13}$ are $(Y^H_{\rm{down}})_{12}$, $\gamma(Y^H_{\rm{down}})_{13}$, 
$\alpha\lambda^{0.5}(Y^C)_{12}$, and $\beta\lambda^{0.5}(Y^C)_{13}$. They are of order $\lambda^5$, and therefore $V_{ub}$ becomes ${\cal O}(\lambda^3)$. In order to obtain
$V_{ub}\sim \lambda^4$, all the leading contributions must vanish. The first two contributions
vanishes if $\langle A\rangle$ is proportional to $Q_{B-L}$,
which plays an important role in solving the doublet-triplet splitting problem. 
This is because $Y^H_{\rm{down}}$ is the same as $Y_u$ and therefore off-diagonal elements
of $Y^H_{\rm{down}}$ are vanishing in the basis in which $Y_u$ is diagonal.
Another important point is that the $U(1)_A$ charge of $C$ is smaller than that of $H$ by one.
This nature forbids the (2,2), (2,3), (3,2) and (3,3) elements (before rotating the basis) of $Y^C$ by the SUSY-zero mechanism\footnote{In our model, the (1,1) entry of $Y^C$ is also forbidden by the discrete symmetry.}.
Therefore, after rotating, $Y^C$ becomes
\begin{equation}
Y^C\sim
\left(\begin{array}{ccc}
1 & \lambda & \lambda^3\\
\lambda & 1 & \lambda^2\\
\lambda^3 & \lambda^2 & 1
\end{array}\right)
\left(\begin{array}{ccc}
0 & \lambda^4 & \lambda^2\\
\lambda^4 & 0 & 0\\
\lambda^2 & 0 & 0
\end{array}\right)
\left(\begin{array}{ccc}
1 & \lambda & \lambda^3\\
\lambda & 1 & \lambda^2\\
\lambda^3 & \lambda^2 & 1
\end{array}\right)
\sim
\left(\begin{array}{ccc}
\lambda^{5} & \lambda^{4} & \lambda^{2}\\
\lambda^{4} & \lambda^{5} & \lambda^{3}\\
\lambda^{2} & \lambda^{3} & \lambda^{5}
\end{array}\right).
\label{eq:YC_small}
\end{equation}
This changes the determination of massless $\bar{\bm{5}}^0_3$ composition (\ref{eq:5bar_solution}) into
\begin{equation}
\alpha\sim\lambda^{2.5}, \ \beta\sim\lambda^{4,5}, \ \gamma\sim\lambda^2,
\end{equation}
so the terms containing $\alpha$ and $\beta$ become smaller.
The above two points are the essential points for $V_{ub}=0$ at the leading order.

Next, we see $(Y_d)_{23}$.
The main contributions are from $\gamma(Y^H_{\rm{down}})_{23}$, $(Y^H_{\rm{down}})_{22}$, $\alpha\lambda^{0.5}(Y^C)_{22}$ and $\beta\lambda^{0.5}(Y^C)_{23}$, but the first contribution
is vanishing if $\langle A\rangle$ is proportional to $Q_{B-L}$\footnote{
If the adjoint Higgs $A$ couples only to the (1,2) (and (2,1)) components as in the model we
are considering, then this condition that  $\langle A\rangle$ is proportional to $Q_{B-L}$
is not necessary, because $Y^H_{\rm{down}}$ is almost $Y_u$ and therefore 
$\gamma(Y^H_{\rm{down}})_{23}$ becomes very small in the basis in which $Y_u$ is diagonal.
}.
The second term is nothing but $(Y_u)_{22}$, so the relation $V_{cb}y_b=y_c$ is realized
when $\alpha$ and $\beta$ become smaller due to the SUSY-zero mechanism as noted above.

How does above discussion changed if we generically include the higher-dimensional operators $\Psi_i\bar{H}H\Psi_jH$ and $\Psi_i\bar{H}H\Psi_jH$?
First, $Y_u$ and $Y^H_{\rm{down}}$ are derived from same expression $\bm{16}_i\bm{16}_j\bm{10}_H$ at $SO(10)$ level, therefore simultaneous diagonalizability of $Y_u$ and $Y^H_{\rm{down}}$ does not affected by the higher-dimensional operators.
Second, the form of $Y^C$ with the SUSY-zero mechanism (\ref{eq:YC_small}) does not changed because adding $\bar{H}H$ ($\bar{h}+h<0$) cannot revive the terms which was forbidden by the SUSY-zero mechanism.
Therefore, the characteristic predictions $V_{ub}\sim\lambda^4$ and $V_{cb}y_b=y_c$ do not affected whether or not we include such higher-dimensional operators.

What becomes obvious in this appendix is that the conditions for $V_{ub}\sim \lambda^4$ and
$V_{cb}y_b=y_c$ can be satisfied in the more general $E_6$ GUTs with family symmetry.
The discrete symmetry for solving the SUSY CP problem is not necessary. One prediction
$V_{ub}\sim \lambda^4$ has been already tested by the B-factory experiments. 
The other prediction $V_{cb}y_b=y_c$ at the GUT scale means that comparably small 
$\tan\beta\equiv \langle H_u\rangle/\langle H_d\rangle $, which can be tested in future
experiments.

\end{document}